\documentclass[12pt]{iopart}
\usepackage{color}
\usepackage{cite}
\begin{document}
\input{epsf}

\title{Energy dependence of strangeness production}

\author{Volker Friese}

\address{ Gesellschaft f\"{u}r Schwerionenforschung, Planckstr. 1, 
64291 Darmstadt, Germany} 

\ead{v.friese@gsi.de}

\author{for the NA49 Collaboration\footnote[7]{for the collaboration
list see \cite{blume}} }


\begin{abstract}

We present a summary of strange particle yields measured by the experiment
NA49 in central Pb+Pb collisions in the beam energy range 20 - 158 $A$GeV. 
The comparison with thermal hadron model results shows that above 
30 $A$GeV, a strangeness undersaturation parameter $\gamma_S$ must be employed 
to describe the data.
This suggests a change in the strangeness production mechanism at this
energy. 
We find that the UrQMD transport model in all cases underpredicts the 
observed yield ratios.

\end{abstract}

\submitto{\JPG}

\section{Introduction}
The experimental data accumulated by several experiments for Pb+Pb 
collisions at top SPS energy (158 $A$GeV) suggest that in the early stage 
of central reactions of this type a deconfined state of matter is 
created \cite{stock,heinz}.
It is commonly believed that at lower beam energies, e.g.~at SIS and AGS,  
such a phase transition does not take place. 
Both statements, although lacking a rigorous proof, motivate a search for 
the onset of deconfinement at energies between top AGS (11.7 $A$GeV) and 
top SPS energy. 
The NA49 collaboration has undertaken such a search by investigating
central Pb+Pb collisions at beam energies between 20 $A$GeV and 158 $A$GeV.

One of the observables possibly signalling a transient deconfined state
in the course of the collision is the relative yield of strange particles.
Originally, it was proposed that strangeness is more easily produced in a 
Quark-Gluon-Plasma than in a purely hadronic system, leading to an 
enhancement of relative strangeness yields in heavy ion collisions in 
comparison to elementary (p+p) reactions once the system underwent a phase 
transition \cite{koch}. 
The interpretation of the observed strange particle yields, however, 
turned out to be more complicated. 
Statistical hadron gas models were successful in describing the measured 
yields, suggesting an equilibrated hadronic system at chemical freeze 
out \cite{stachel}. 
In such models, the strangeness enhancement in Pb+Pb collisions is a
consequence of the increased system volume, relaxing the constraints 
imposed by flavour conservation which result in a suppression of 
strangeness in small systems \cite{rafelski,tounsi}. 
It is still under debate whether an equilibrium description also holds 
for strange particles, i.e.~whether a strangeness undersaturation factor 
$\gamma_S$ is needed to describe the data \cite{becattini}. 
This factor could be interpreted as the influence of a prehadronic stage 
onto the available hadronic phase space \cite{csernai}.
More data at various beam energies and collision system sizes will help to 
answer such questions.

As an early result of the NA49 energy scan programme, a sharp maximum in 
the $K^+ / \pi^+$ excitation function at 30 $A$GeV ($\sqrt{s_{NN}} = 7.6$ GeV)
was reported in the last 
conference of this series \cite{friesesqm03}, consistent with the 
assumption of the onset of deconfinement at that energy \cite{gazdzicki}. 
Meanwhile, the energy scan has been completed by taking data at the lowest 
SPS energy (20 $A$GeV). 
In addition, the available data have been analysed with respect to the 
production of other strangeness carriers. 
We present a summary of results obtained in the context of the energy scan 
programme by NA49 so far and compare them to data obtained at AGS and RHIC 
and to predictions of statistical and microscopic models. We concentrate 
on particle yields; a discussion of the kinematic distributions can be 
found in \cite{blume}. 
We choose UrQMD \cite{urqmd13,urqmd21} as a representative of 
hadronic transport models.
As thermal models we employ versions with and without the assumption of 
strangeness saturation. 
The former, in the following labelled as Hadron Gas Model (HGM), has 
predictive power by parametrising the variation of $T$ and $\mu_B$ with 
collision energy \cite{hgm}. 
The latter, denoted by Statistical Hadronisation Model (SHM), fits the 
observed yields using $\gamma_S$ to describe the deviation of strange particle 
yields from full equilibrium \cite{shm}.

In the context of hadron gas models, it is debated whether particle 
yields around midrapidity or fully integrated over rapidity should be 
considered.
In the former case, one stays away from the fragmentation regions where 
the properties of the system may be different from those in the central 
region. 
However, the data obtained at AGS, SPS and even RHIC reveal nearly 
Gaussian rapidity distributions for the produced particles \cite{roehrich}, 
showing no sign of a clear separation of central and fragmentation regions. 
Moreover, by using phase space integrated yields, the fit results are 
independent of the kinematical behaviour of the system as e.g.~flow. 
By studying integrated yields, as we choose to do in this report, one of 
course addresses the average properties of the system (the ``Equivalent 
Global Cluster'' \cite{shm}) and is not sensitive to possible variations 
of thermal parameters with rapidity. 

\section{Experiment and data analysis}

The NA49 experiment \cite{na49nim} is a large acceptance hadron spectrometer 
operated at CERN-SPS with external heavy ion beams. 
It consists of two superconducting dipole magnets and four large-volume TPCs 
for tracking and momentum measurement. 
The identification of charged particles is achieved by the measurement of 
specific energy loss in the TPC gas and, in a restricted acceptance region 
around midrapidity, by time-of-flight measurement in two scintillator 
arrays located at 14 m from the target. 
The event centrality is determined from the energy deposited by the beam 
spectators in a zero-degree calorimeter. 
For the data sets presented here, the online centrality trigger selected 
the 7 $\%$ most central events except at 158 $A$GeV, where the centrality 
was 5 $\%$ for kaons and $\phi$, 10 $\%$ for $\Lambda$ and $\Xi$ and 23.5 $\%$ 
for $\Omega$.
The data at lower energies (20, 30, 40 and 80 $A$GeV, corresponding to
$\sqrt{s_{NN}} = $ 6.3, 7.6, 8.8 and 12.3 GeV, respectively)
were taken in the years 1999, 2000 and 2002, 
the data at top SPS energy ($\sqrt{s_{NN}} = 17.3$ GeV) in 1996 and 2000.

For the identification of charged particles, TOF is used at momenta below
2.5 GeV to separate pions, kaons and protons. 
Between 2.5 and 10 GeV, but still at midrapidity, the combined TOF and 
$dE/dx$ measurements still allow a clean separation almost on a 
track-by-track basis. 
At higher momenta, outside the TOF acceptance, the contributions of the 
particle species to the energy loss spectrum are resolved on a statistical 
basis \cite{friesesqm03}.

Hyperons are identified in the invariant-mass spectrum of their decay
daughters, using the decay topology to suppress the combinatorial
background. 
Resonance signals are also extracted from the invariant-mass spectrum, 
which is constructed after appropriate PID cuts on their decay products.

The strange particle yields presented in the next sections will be
normalised to the total pion yield
$ \langle \pi \rangle = 1.5 \times ( \langle \pi^+ \rangle + \langle \pi^- \rangle ) $. 
The $\pi^-$ yield was derived by subtracting the contributions of $K^-$,
secondary interactions and feeddown from weak decays from the negatively 
charged hadron yield measured by NA49. 
From this, the $\pi^+$ yield was calculated using the $\pi^+ / \pi^-$ ratio that 
was determined in the limited TOF acceptance. 
The method is described in detail in \cite{na49energy}. 
It should be noted that the pion yields at AGS and SPS are corrected for 
the contributions of weak decays. 
This does not hold for the RHIC data, where this contribution is estimated 
to be 4 $\%$ \cite{roehrich}.

\section{Results}

\subsection{Kaons}
The transverse mass spectra for $K^+$ and $K^-$, measured at midrapidity in 
the TOF acceptance for the five beam energies show, within the experimental 
uncertainties, no deviations from an exponential shape. 
The slope parameters obtained by an exponential fit are similar for both 
particle types and do not vary significantly with beam 
energy \cite{marekqm04}. 
The acceptance covers more than 95 $\%$ of the total yield. 
The rapidity distributions, obtained by integration of the transverse 
spectra in different rapidity bins, can be described by the sum of two 
Gaussians displaced symmetrically around midrapidity \cite{blume}. 
For the $K^-$, a single Gaussian fit is equally satisfying. 
At larger beam energies, the TOF points at midrapidity agree well with the 
results obtained by the $dE/dx$ analysis, showing the consistency of these 
two identification methods. 
At the lower energies, there is no overlap due to the lower momentum cutoff
for the $dE/dx$ method. 
The width of the rapidity distributions increases linearly with the beam 
rapidity.

\begin{figure} 
\begin{center}
\epsfxsize=16cm
\epsfbox{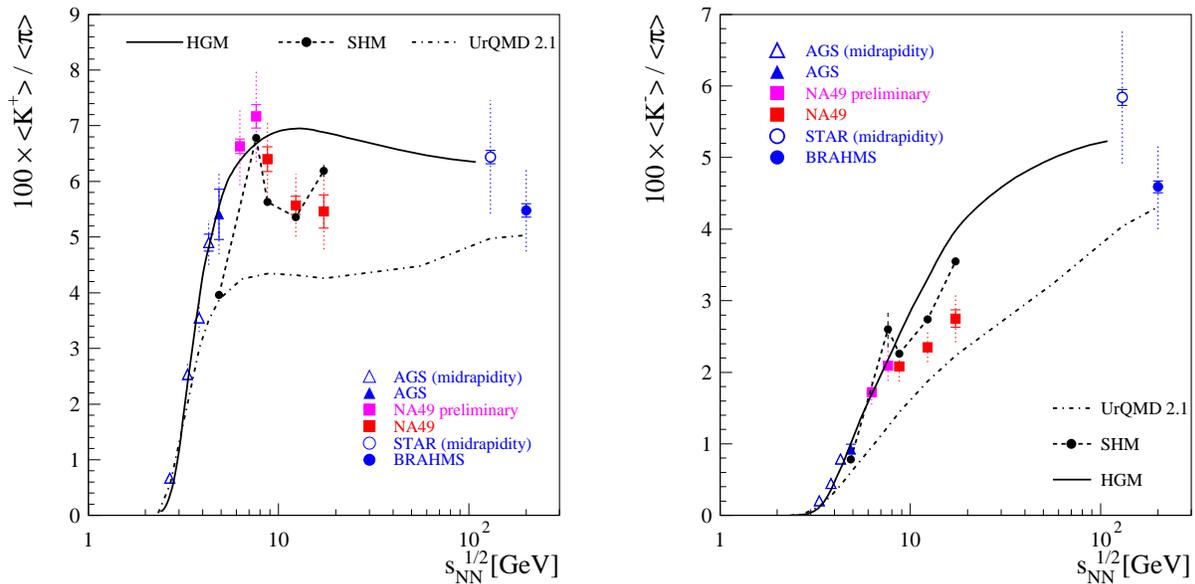}
\end{center}
\caption{\label{fig:ka2pi} 
The $K^+ / \pi$ (left) and $K^- / \pi$ (right) ratios  in central Pb+Pb (Au+Au) 
collisions as functions of CM energy.
Open symbols denote measurements around midrapidity, closed symbols those
in full phase space. 
The full error bars give the statistical, the dashed ones the sum of 
statistical and systematic errors. The NA49 data points correspond to
20, 30, 40, 80 and 158 $A$GeV beam energy, respectively. 
The full line shows the predictions of the HGM, the dots the fit results of 
the SHM, which are connected by the dashed line to guide the eye. The error
of the SHM is derived from the difference between fits A and B in \cite{shm}.
The dashed-dotted curve represents the results of UrQMD 2.1.
Data are NA49 preliminary or taken from 
\cite{na49energy,e895pion,e802pion,starpion,brahms,e866kap,e866kam,e802kaon,
starkaon}
}
\end{figure}

By integration of the rapidity distribution, total yields are obtained. 
The energy dependence of the $\langle K^+ \rangle / \langle\pi \rangle$ ratio is shown in figure 
\ref{fig:ka2pi} (left) together with the model results. 
The newly obtained data point at 20 $A$GeV ($\sqrt{s_{NN}} = 6.3$ GeV)
confirms the sharp peak 
observed in this variable as reported earlier \cite{friesesqm03}. 
The HGM, assuming a smooth variation of the $T$ and $\mu_B$ parameters with 
collision energy, is not able to reproduce this feature, although it 
predicts a broad maximum at low SPS energies. 
In particular, the relative $K^+$ yield at higher SPS energies is 
overpredicted. 
The SHM, having $\gamma_S$ as an additional parameter, is better able to account 
for the data. 
UrQMD predicts a small increase in this ratio, but underestimates its value 
at all SPS energies.

In contrast to the $K^+$, the $\langle K^- \rangle / \langle\pi \rangle$ ratio, shown in figure 
\ref{fig:ka2pi} (right), exhibits no sharp structure but a smooth 
evolution with collision energy, with possibly a small deviation around 
30 $A$GeV ($\sqrt{s_{NN}} = 7.6$ GeV). 
The comparison with the models shows again that HGM overpredicts the data 
at higher SPS energies. 
In the SHM fit, the peak in $K^+/\pi$ drives $\gamma_S$ almost to unity resulting 
in a similar feature in $K^-/\pi$, which is not as pronounced in the data. 
The UrQMD results are systematically too low, but closer to the data than 
in the case of $K^+/\pi$. 

\subsection{$\Lambda$ and $\bar{\Lambda}$}

The $\Lambda$ and $\bar{\Lambda}$ measurements of NA49 \cite{na49lambda} have
been extended to the lower SPS energies. 
While the $\bar{\Lambda}$ rapidity distribution is approximately Gaussian at all 
energies, the $\Lambda$ distribution visibly flattens above 40 $A$GeV \cite{blume}. 
At top SPS energy, this leads to a larger error in the extrapolation to 
full phase space. 

\begin{figure} 
\begin{center}
\epsfxsize16cm
\epsfbox{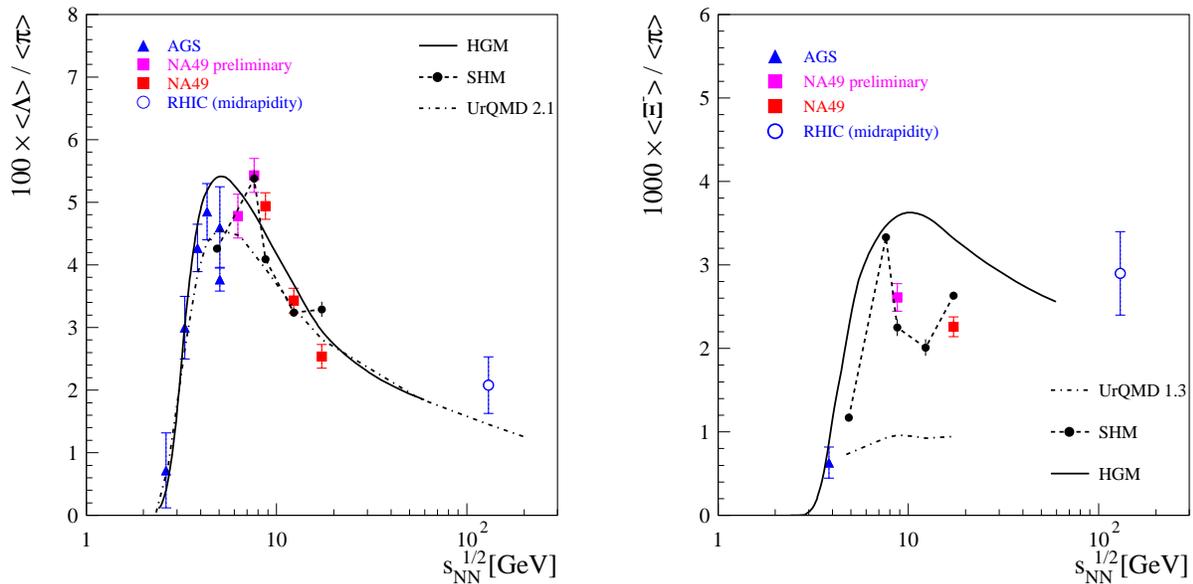}
\end{center}
\caption{\label{fig:lamxi2pi} 
Same as figure \ref{fig:ka2pi}, but for the $\Lambda/ \pi$ (left) and $\Xi^- / \pi$ 
(right) ratios. 
Only statistical errors are shown.
Data are NA49 preliminary or taken from 
\cite{e895pion,e802pion,na49energy,starpion,brahms,e895lambda,e896lambda,
na49lambda,starlambda,e895xi,na49xi,starxi}
}

\end{figure}

Normalised to the pion yield, the $\Lambda$ excitation function shows a sharp
maximum around 30 $A$GeV ($\sqrt{s_{NN}} = 7.6$), 
similar to the $K^+$, although AGS data in this
case are not conclusive (figure \ref{fig:lamxi2pi} (left)). 
As $\Lambda$ and $K^+$ are the bulk carriers of  $s$ and $\bar{s}$ quarks, 
respectively, and the net strangeness has to vanish, this proves the 
consistency of the NA49 measurements, the identification methods for kaons 
and $\Lambda$ being completely different.
Although predicting the maximum of $\langle \Lambda \rangle / \langle \pi \rangle$ to be at top AGS energy, 
the overall agreement of the HGM with the data is reasonable. 
A still better description is  provided by the SHM while UrQMD is close to 
the measurements at AGS and top SPS but misses the peak at lower SPS energies.

\subsection{$\Xi^-$}

\begin{figure} 
\begin{center}
\epsfxsize16cm
\epsfbox{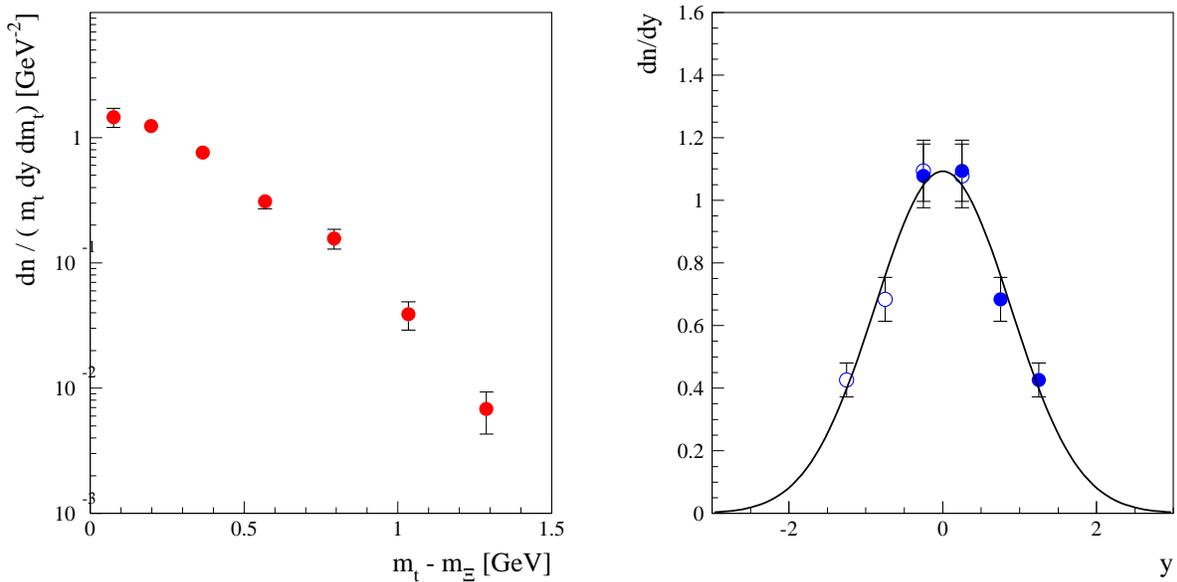}
\end{center}
\caption{\label{fig:xi40} 
Transverse mass (left) and rapidity (right) spectra of the $\Xi^-$
in central Pb+Pb collisions at 40 $A$GeV.
Open points are reflected at midrapidity.
The full line shows a Gaussian fit.}

\end{figure}

Multiply strange particles, although carrying only a small fraction of
the total strangeness, can be expected to discriminate better between
statistical models with and without $\gamma_S$ since this factor will enter
with a higher power into the predicted yields than for kaons and $\Lambda$.
Naturally, data for such rare particles are scarce.  
In addition to the cascade measurement at top SPS energy \cite{na49xi}, 
the $\Xi^-$ signal has been analysed by NA49 at 40 $A$GeV \cite{meurerqm04}. 
The $\Xi^-$ transverse mass and rapidity distributions are shown in figure 
\ref{fig:xi40}.
Once more, a single Gaussian gives a satisfactory fit to the rapidity 
spectrum, indicating the diminishing influence of baryon density with 
increasing strange quark content. 
As shown in figure \ref{fig:lamxi2pi} (right), the integrated $\Xi^- / \pi$ 
ratio is slightly higher than at 158 $A$GeV. 
Both data points are overpredicted by the HGM while again, the SHM is 
closer to the measurements. 
UrQMD by far cannot account for the measured yields.

The NA49 cascade analysis is still in progress; promising signals are seen 
at all beam energies. 
The excitation function thus can be expected to be better mapped in the 
near future.

\subsection{$\Omega^-$ and $\bar{\Omega}^+$}

\begin{figure} 
\begin{center}
\epsfxsize=16cm
\epsfbox{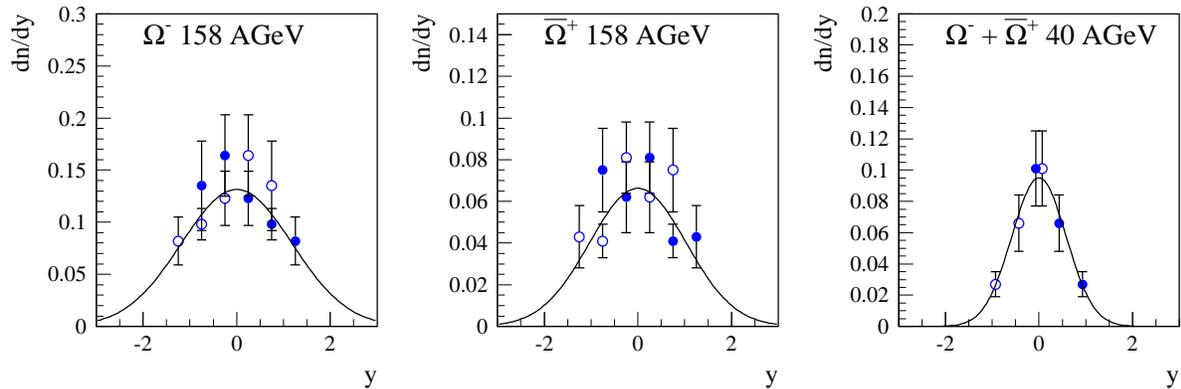}
\end{center}
\caption{\label{fig:omegarap} 
Rapidity distributions for $\Omega^-$ and $\bar{\Omega}^+$ at 158 $A$GeV and for 
$\Omega^- + \bar{\Omega}^+ $ at 40 $A$GeV. Open points are reflected at midrapidity.
The full lines show Gaussian fits to the distributions.}
\end{figure}

For the $\Omega$ measurement at 158 $A$GeV, the high statistics run of 2000
with a 23.5 $\%$ online centrality trigger had to be exploited. 
At 40 $A$GeV with considerably less statistics, only the combined 
$\Omega^-+\bar{\Omega}^+$ signal could be studied. 
Figure \ref{fig:omegarap} shows the rapidity distributions for both 
energies. 
Within the considerable uncertainties, a Gaussian function gives a good 
description. 
As shown in figure \ref{fig:omphi2pi} (left), the yield ratio 
$(\Omega^- + \bar{\Omega}^+) / \pi$ shows a steady increase from 40 $A$GeV 
($\sqrt{s_{NN}} = 8.8$ GeV) over top 
SPS energy to RHIC. 
Both flavours of statistical models manage to describe the data within the 
experimental errors. 
The small difference between the model results is surprising
since naively, one would expect them to differ noticeably, the factor 
$\gamma_S$ entering to the third power. 
As already observed in the case of cascades, the relative $\Omega$ yield is 
underestimated strongly by UrQMD.


\subsection{$\phi$ mesons}
In the context of strangeness production, the $\phi$ meson is of special
interest because of its strangeness neutrality as a hadron. 
It should therefore not be sensitive to canonical effects if strangeness 
production is assumed to happen in a hadronic scenario. 
It is also not clear whether a possible undersaturation factor $\gamma_S$ should 
be applied or not. 

\begin{figure} 
\begin{center}
\epsfxsize=16cm
\epsfbox{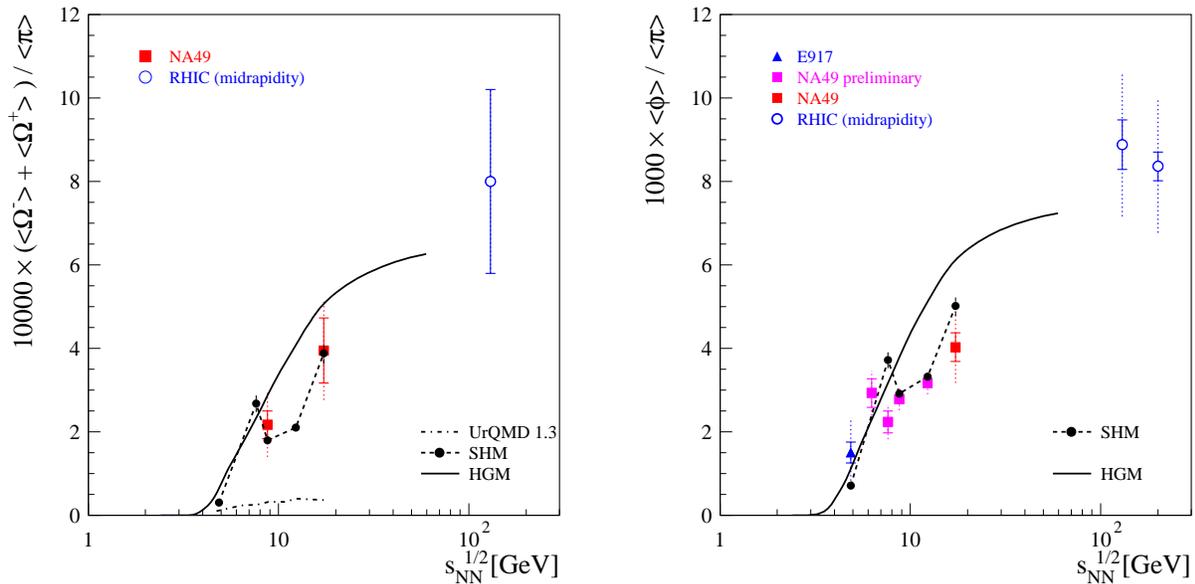}
\end{center}
\caption{\label{fig:omphi2pi} 
Same as figure \ref{fig:ka2pi}, but for the $\Omega / \pi $ (left) and 
$\phi / \pi$ (right) ratios.
Data are NA49 preliminary or taken from 
\cite{e802pion,na49energy,starpion,brahms,staromega,e917phi,na49phi,
starphi130,starphi200}
}
\end{figure}

The NA49 $\phi$ measurement at 158 $A$GeV \cite{na49phi} has been extended to
all available energies. 
Figure \ref{fig:omphi2pi} (right) shows the normalised $\phi$ yield as 
function of collision energy. 
The excitation function is similar to that of the $K^-$, showing an overall
increase with beam energy.
A careful study of the systematic errors is required to assess the 
significance of the apparent local minimum at 30 $A$GeV 
($\sqrt{s_{NN}} = 7.6$ GeV). 
As in the case of $K^-$, the HGM overpredicts the yield at higher SPS 
energies, while the SHM fits are able to reproduce the data with the 
exception of 30 $A$GeV. 
At that energy, the model fit is driven to a maximum as in the case
of the $K^+$, but the contrary is seen in the data.
The $\phi$ yield in UrQMD is not straightforwardly defined since $\phi$ decays 
happen not only after chemical freezeout but in all stages of the
simulated collision, and the rescattering of the decay products will affect
the number of detectable $\phi$.

\section{Discussion} 

Summarising the comparison of measured particle yields with the models,
we find as a general trend that the Hadron Gas Model reproduces the
normalised yields at lower collision energies but starts to overpredict
them from around 30 $A$GeV on, with the exception of $\Lambda$ and $\Omega$. 
At RHIC energies, the measured data are again well reproduced. 
Apparently, in the SPS energy range the introduction of a 
strangeness suppression factor $\gamma_S$ gives a much better description
of the data. 
At 30 $A$GeV both model types coincide since here, strangeness seems to be 
saturated, resulting in $\gamma_S$ close to unity. For the other energies,
the best fit yields values of 0.7 - 0.8 for this parameter.
The deviation from the saturation model suggests a basic change
in the strangeness production mechanism at around 30 $A$GeV beam energy.
Interestingly, the $\phi$ meson seems to call for $\gamma_S$ as well as the other 
strange hadrons, indicating that $\gamma_S$ reflects conditions of a prehadronic
stage of the collision. 
For the interpretation of the $\Omega$ yields, it is unclear why both model 
types give similar results, contrary to the expectation.

A feature of the HGM is a maximum in the relative strangeness content 
as a consequence of the interplay between decreasing baryochemical
potential and increasing temperature. 
This maximum, however, is relatively broad and cannot account for the 
sharp structures seen in the excitation functions of the bulk strangeness 
carriers. 
Moreover, its position is shifted towards higher collision energies with 
increasing strangeness content and is different for mesons and baryons 
\cite{hgm}.
In this context, the completion of the $\Xi$ excitation function can be
expected to have additional discriminating power. 

In all cases, the UrQMD model underestimates the relative strangeness
yields. 
The discrepancy increases with the strange quark content of the
particle; it is largest for the $\Omega$ with three strange valence quarks.
For the $K / \pi$ ratio, it can be shown that the discrepancy stems from
an overprediction of pions rather than a underestimation of kaons,
but this cannot account for the strong underestimation of the multiply
strange hadron yields. 
Apparently, the UrQMD transport model misses some basic features of the 
underlying strangeness production process.

Considering the total strangeness production in the collision, the
sharp peak in the $K^+ / \pi $ ratio is confirmed by a similar feature
in the $\Lambda / \pi$ ratio. 
It appears doubtful whether such a sharp structure can be explained by a 
continuous variation of parameters within hadronic models. 
So far, the only model to predict a sharp maximum in the relative
strangeness content is the Statistical Model of 
the Early Stage \cite{gazdzicki}, assuming the transition to a deconfined 
state to set in at 30 $A$GeV beam energy. While this model does not
predict the distribution of the produced strangeness among the hadron
species, it is obvious that in the high net baryon density environment
created at this energy, the $s$ quarks will appear in $\Lambda$ baryons
rather than in $K^-$, thus explaining the absence of a similar sharp 
structure in the $K^- / \pi$ ratio.
We conclude that the data collected by NA49 favour such a scenario.

\ack
This work was supported by the US Department of Energy
Grant DE-FG03-97ER41020/A000,
the Bundesministerium f{\"{u}r Bildung und Forschung, Germany, 
the Polish State Committee for Scientific Research 
(2 P03B 130 23, SPB/CERN/P-03/Dz 446/2002-2004, 2 P03B 04123), 
the Hungarian Scientific Research Foundation (T032648, T032293, T043514),
the Hungarian National Science Foundation, OTKA, (F034707),
the Polish-German Foundation, and the Korea Research Foundation Grant 
(KRF-2003-070-C00015).



\end{document}